\documentclass[12pt]{article}

\usepackage{psfig}
{\end{list}}
\newcounter{enumct}

\newcommand{\captive}[1]{\rule{5mm}{0mm}%
\begin{minipage}{150mm}\caption[small]{#1}\end{minipage}}

\def\APP{{\em Acta Physica Polonica} }
\def\NIMA{{\em Nucl. Instrum. Methods} A}
\def\NPB{{\em Nucl. Phys.} B}
\def\PLB{{\em Phys. Lett.}  B}

\def\PRD{{\em Phys. Rev.} D}
\def\ZPC{{\em Z. Phys.} C}

\def\EPJC{{\em Eur. Phys. Journal} C}
\def\ARNS{{Annu. Rev. Nucl. Sci.}}

\def\NIMA{{\em Nucl. Instrum. Methods} A}

\def\kbn{Polish Committee for Scientific Research 2P03B01414}

\def\ra{\rightarrow}

\def\be{\begin{equation}}
\def\ee{\end{equation}}
\def\bea{\begin{eqnarray}}
\def\eea{\end{eqnarray}}

\begin{document}
 
\sloppy

\begin{flushright}
IFT 98/18\\[1.5ex]
{\large \bf hep-ph/9812536} \\
\end{flushright}

\begin{center}
{\LARGE\bf Constraints on the Higgs sector from processes involving photons}
\footnote{ Presented at the Workshop on Photon Interactions and the 
Photon Structure, Lund, September 10-13, 1998.} 
\\[4mm]
{\Large Maria Krawczyk} \\[3mm]
{\it Institute of Theoretical Physics,}\\[1mm]
{\it  Warsaw University, Warsaw, Poland}\\[1mm]
{\it E-mail: krawczyk@fuw.edu.pl}\\[20mm]

{\bf Abstract}\\[1mm]
\begin{minipage}[t]{140mm}
The Higgs sector of the Standard Model and of the Two Higgs Doublet 
Extensions of SM, MSSM and the general 2HDM, can be tested in
 processes involving photons.
A short review of the corresponding results is presented.
\end{minipage}\\[5mm]

\rule{160mm}{0.4mm}

\end{center}

\section{Introduction}
The Higgs boson is still a missing element of the 
Standard Model (SM).
The direct searches performed at LEP lead to the 95\% C.L. lower limit
 \cite{desch} $$ M_{Higgs}\ge 89.8 \rm ~~GeV.$$
The indirect analyses of the all electroweak data, 
where the quantum effects due to the 
other fundamental particles are taken into account, 
 prefer the light SM Higgs boson  (with mass below 262 GeV 
at 95 \% C.L., \cite{hollik}).
The requirements of vacuum stability and the validity
of perturbative theory up to the unification scale 
give   the mass of the SM Higgs boson to be approximately 
between 130 and 180 GeV \cite{hollik,limth}. 

In the Two Higgs Doublet Extensions of the SM, 
like Minimal Supersymmetric
Standard Model (MSSM), more Higgs bosons are expected to exist:
neutral scalars $h$ and $H$ (with mass $M_H\ge M_h$), 
a pseudoscalar (more properly a CP-odd particle) $A$
and charged scalars $H^{\pm}$ \cite{hunter}. Such models
are in addition to masses characterized by two parameters (assuming
 the CP conservation): 
a ratio of vacuum expectation values for two scalar fields
$\tan \beta=v_2/v_1$ and $\alpha$, describing the mixing 
between two neutral scalar Higgs particles $h$ and $H$.
These parameters govern the corresponding couplings of Higgs bosons to 
 themselves (here the extra parameter, $\lambda_5$, appears)  
and to gauge bosons and fermions. 
 Some of the Higgs couplings can be enhanced, some of them suppressed 
depending  on values of  parameters. 
In the so called Model II,  one of  Higgs doublets
couples only to ``up'' components of the isodoublets, 
the other to the ``down''
components - this way the FCNC are naturally avoided
(at the tree level). Then   
for large $\tan \beta$  couplings of $h$ and $A$ to ``down-type''
 quarks and charged leptons are highly enhanced while those 
 to ``up-type'' quarks suppressed. Couplings to gauge bosons
 of the $h$ and $H$ are proportional to 
 $\sin(\beta-\alpha)$ and $\cos(\beta-\alpha)$, respectively.
The $A$ boson does not couple to $Z/W$ bosons.

 In the MSSM there appear relations between parameters of the model:
$\alpha$, $\beta$ and Higgs masses, leaving only two of them as  
independent
 parameters at the tree level.  Therefore  tide constraints
 on the Higgs boson masses are expected (with a 
surprisingly weak dependence on the sector of supersymmetric particles). 

The non-supersymmetric version of the Model II, denoted here as 2HDM, 
 has a  Higgs sector the same 
as  MSSM but the relations between parameters imposed by the supersymmetry 
are missing.
Therefore each parameter has to be constrained independently.
 Vacuum stability analysis suggests that this model can not be valid up
 to the unification scale \cite{nie}. This is exactly   
what is expected 
since  the considered model 
  can be treated as a low energy realization of the more fundamental theory.

We will see below (Sec.2) how  much 
phenomenological consequences differ in these two approaches.
The most striking difference is that
much lighter Higgs bosons are allowed by the same data in 2HDM 
than in MSSM case.

The processes involving photons can test the Higgs sector of the SM and
of its extensions. In my talk I will focus on the search for light neutral 
Higgs bosons in the framework of 2HDM at present (Sec. 3.1) and future
experiments (Secs.3.2, 4 and 5). 
\section{Present limits on a non-minimal Higgs bosons}

Not only in SM but also in  MSSM  neutral Higgs boson $h$ is 
 expected to be light, with mass below 135 GeV
(\cite{hollik,carena}). On the other hand
the present lower 95\% C.L. limits   from direct searches at LEP 
(for the CM energies up to 183 GeV)
are   as follows \cite{desch}
$$    M_h\ge 77 {\rm ~GeV~~and}~~M_A\ge 78 {\rm ~GeV~~} $$
for 
$$\tan \beta \le 0.8 {\rm ~~or} ~~\tan \beta \ge 2.1.$$
In Fig. 1a,b the OPAL results for the pseudoscalar and scalar 
based on  LEP data (with the CM energy 183 GeV) are shown
for $\tan \beta$ as a function of $M_A$ and $M_h$.
 For the allowed ($M_h,M_A$) region, see Fig.2a.

 In 2HDM 
even a lighter neutral Higgs particle is allowed by the present data,
provided the other Higgs particles are heavy enough, 
$M_h+M_A\ge 50 $ GeV, see Fig.2b.
 Two scenarios are worth to be studied here:\\

$\ra$ with a (very) light scalar $h$\\

$\ra$ with a (very) light pseudoscalar $A$.\\

\begin{figure}
\begin{center}
\hskip -3.5in
\mbox{\psfig{file=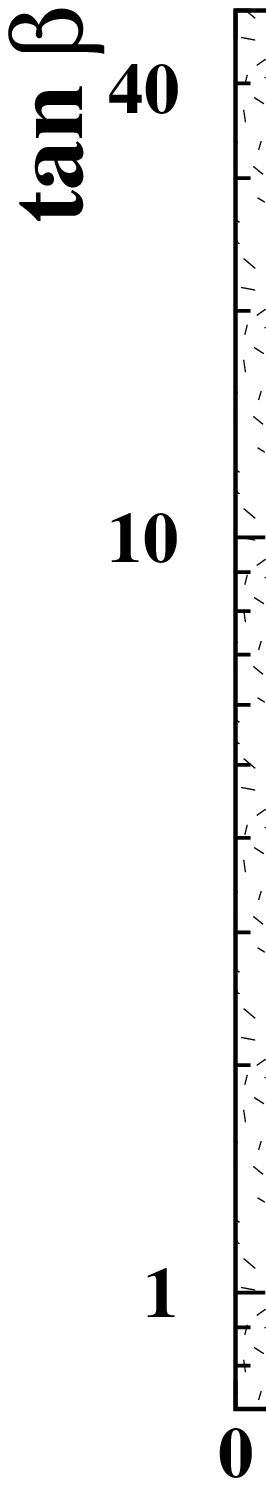,width=79mm}}
\hskip -3.5in
\mbox{\psfig{file=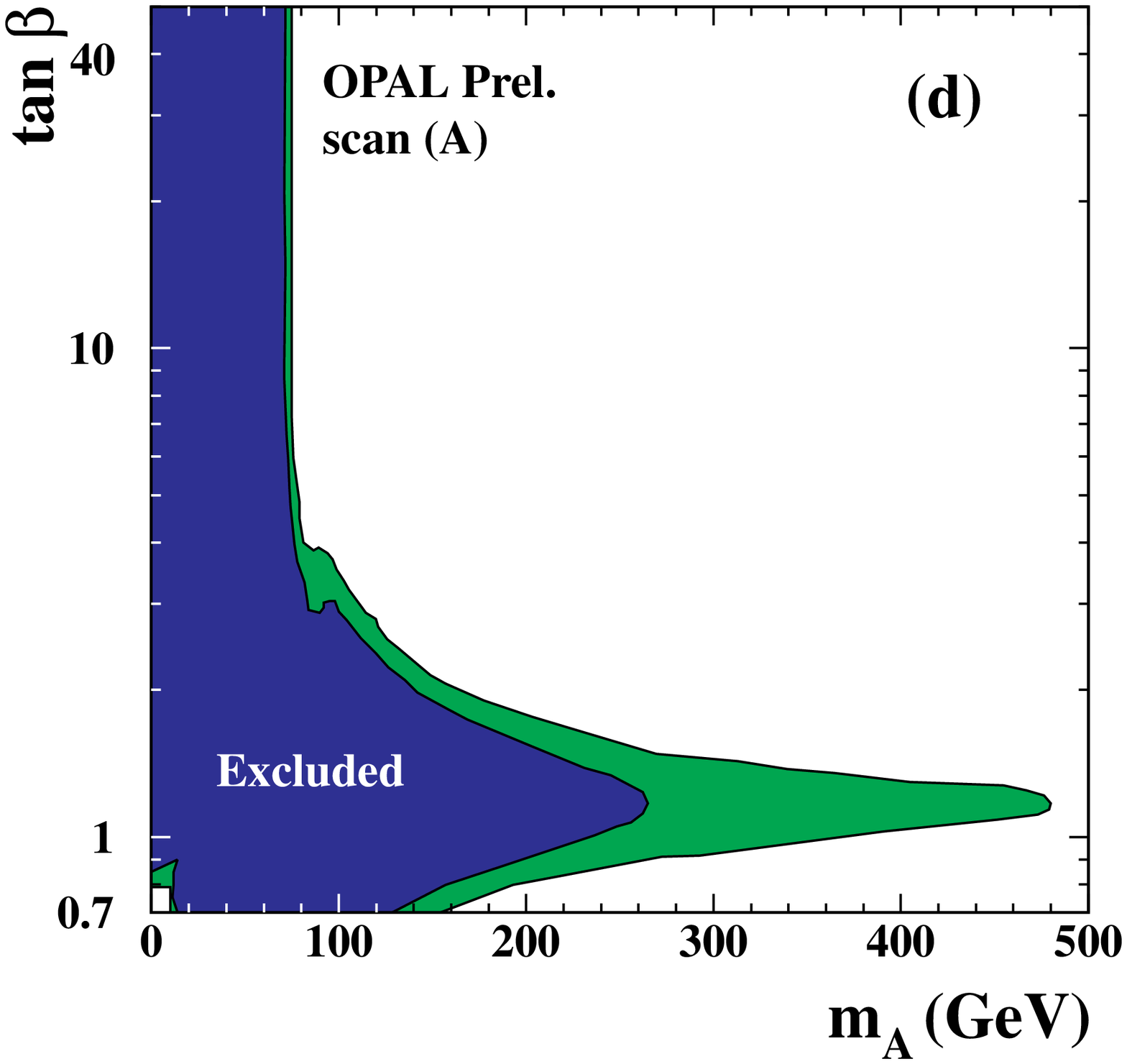,width=79mm}}
\end{center}
\captive{
a) The OPAL  limits based on the LEP data at the CM energy 183 GeV
for a) $\tan \beta$ versus mass $M_A$, b)
$\tan \beta$ versus mass $M_h$
\cite{opal361}. 
\label{figure}}
\end{figure}
\begin{figure}
\begin{center}
\hskip 0.in
\mbox{\psfig{file=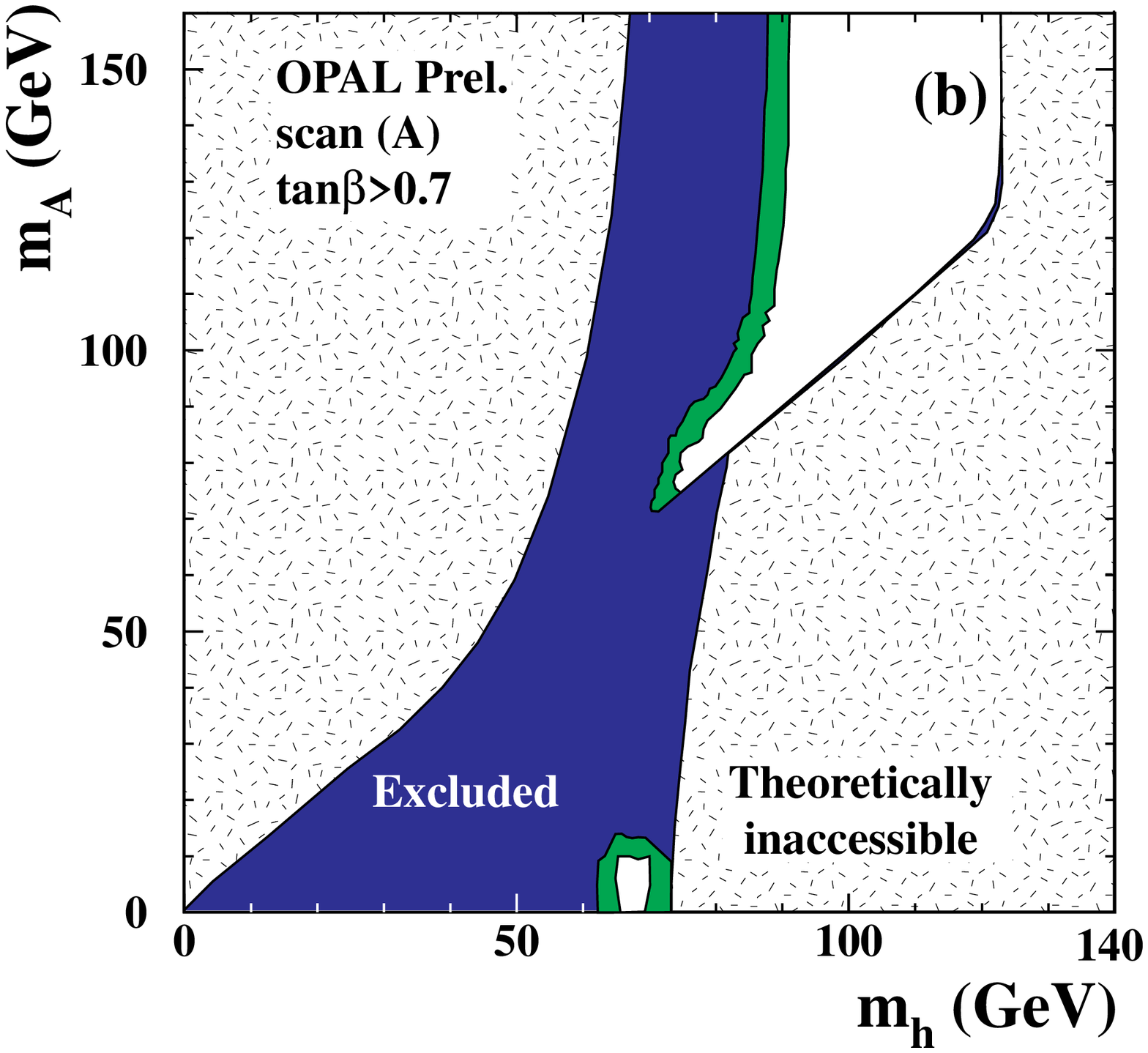,width=79mm}}
\hskip 0.in
\mbox{\psfig{file=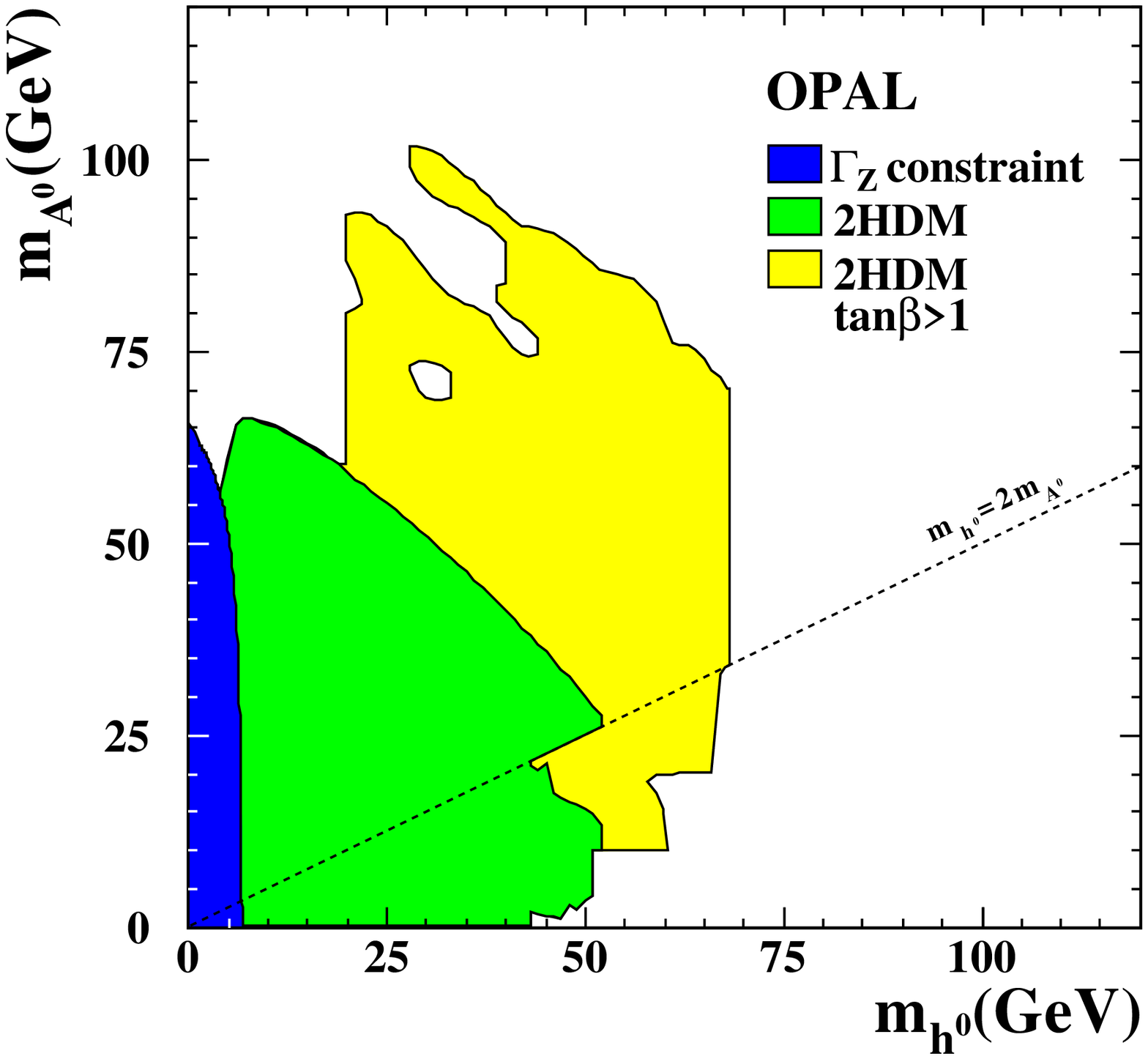,width=79mm}}
\end{center}
\captive{a) The  limits for  Higgs boson masses  $M_A$ versus  $M_h$ from OPAL
analysis obtained in a) MSSM \cite{opal361}, b) 2HDM \cite{opal366}. 
\label{figure}}
\end{figure}
The strength of the pseudoscalar coupling to fermions, $\tan \beta$, 
in context of 2HDM was studied at LEP I by ALEPH group in the Yukawa process
$e^+e^-\ra Z \ra {\bar f} f A $ \cite{alephyu}. 
These results are plotted in Fig. 3 (denoted ``Yukawa'') to be compared to
analogous results  presented in Fig. 1a in context of MSSM.

For charged Higgs boson mass  there are constraints 
from the direct search performed at LEP,
at  the CM energy up to 183 GeV. The 
 95\% C.L. limit is
$$M_{H^{\pm}}\ge 56-59 {\rm ~GeV}$$
for four LEP experiments \cite{desch}. This limit should hold
both for MSSM and 2HDM, since
the $ZH^+H^-$ coupling  
responsible for a charged Higgs bosons production at LEP 
does not depend on  specific parameters of the Model II.\\
The undirect  limit on the mass of a charged Higgs boson
arises from the process  $b\ra s \gamma$. 
This process is mediated by loops  and therefore it is a 
probe of the Standard Model and of its  possible extensions.
In context of 2HDM one gets $M_{H^\pm}$ to be above 165 GeV \cite{greub}
for $\tan \beta$ larger than 2.
(The analysis in the MSSM is more involved and will not be discussed here.)

The additional constraints on the Higgs bosons in the 2HDM 
arise from the data on the $g-2$ for muon \cite{g2,mkwar}, see Fig. 3.
\section{The process $Z\ra h(A) + \gamma$}
\subsection{LEP I} 
\begin{figure}
\begin{center}
\mbox{\psfig{file=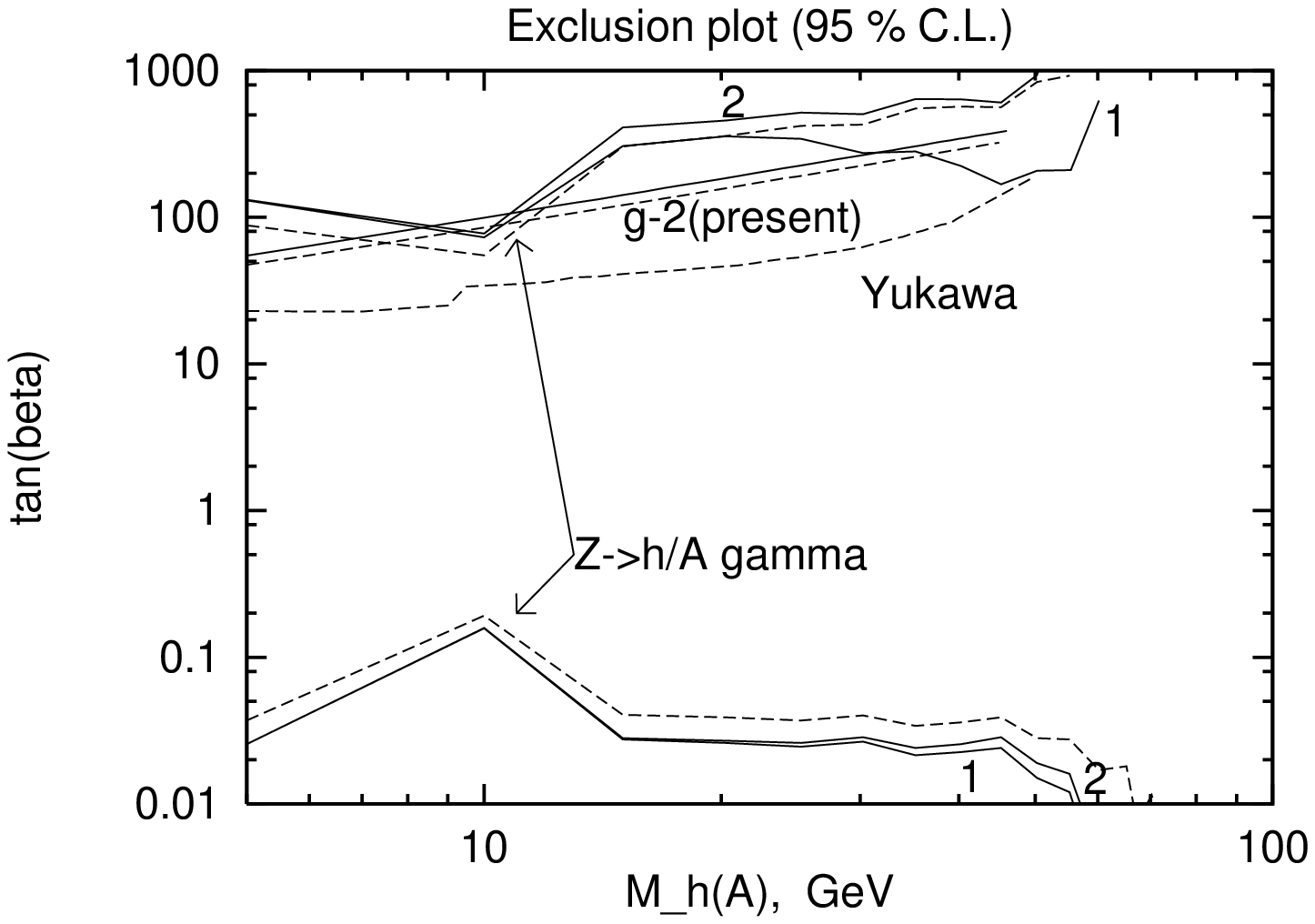,width=120mm}}
\end{center}
\captive{
 The present limits for $\tan \beta$ versus mass $M_h$ (solid lines)
 or $M_A$ (dashed lines) from
analysis of the $Z\ra h(A) +\gamma$ process at LEP I,
compared to constraints from  the $g$-2 for muon data \cite{g2}
 and the Yukawa process $e^+e^- \ra {\bar f} f A $ at LEP I \cite{alephyu}.
The regions above the upper and below the lower curves are excluded.
For the scalar production  experimental limits on $\sin(\beta-\alpha)$
from L3 \cite{sinbal3} are included and two masses of the
 charged Higgs boson are assumed: 1) 54.5 GeV and 2) 300 GeV.  From \cite{mkz}.
\label{figure}}
\end{figure}

With the above  limits in mind we can discuss now  results from
the analysis of  the $Z\ra h(A) + \gamma$ process, measurements of which
were performed recently at the $Z$-resonance by all four LEP experiments.
The measured branching ratio  is of order 10$^{-6}$ to 10$^{-5}$
 \cite{brzhag,mkz}.
In the SM the scalar production  is due to the $W$
 and the fermions loop contributions 
(with a strong domination of the $W$-loop).
 The data
are  well above the SM prediction. 

In the 2HDM the $Z\ra h+\gamma$
 proceeds  via   mentioned  loops with $W$ and fermions,
 now with   different couplings
depending on the parameters $\alpha$ and $\beta$,
and in addition via  a charged Higgs boson loop. 
For the pseudoscalar production        
only fermions contribute.
The results are given Fig.3  in form of  the 95 \% C.L. exclusion plot
for $\tan \beta$ versus mass of the $h$ or $A$.
The comparison is made with the results based on the Yukawa process
$e^+e^-\ra f {\bar f} A$, and the muon anomalous magnetic moment data.

\begin{figure}
\begin{center}
\mbox{\psfig{file=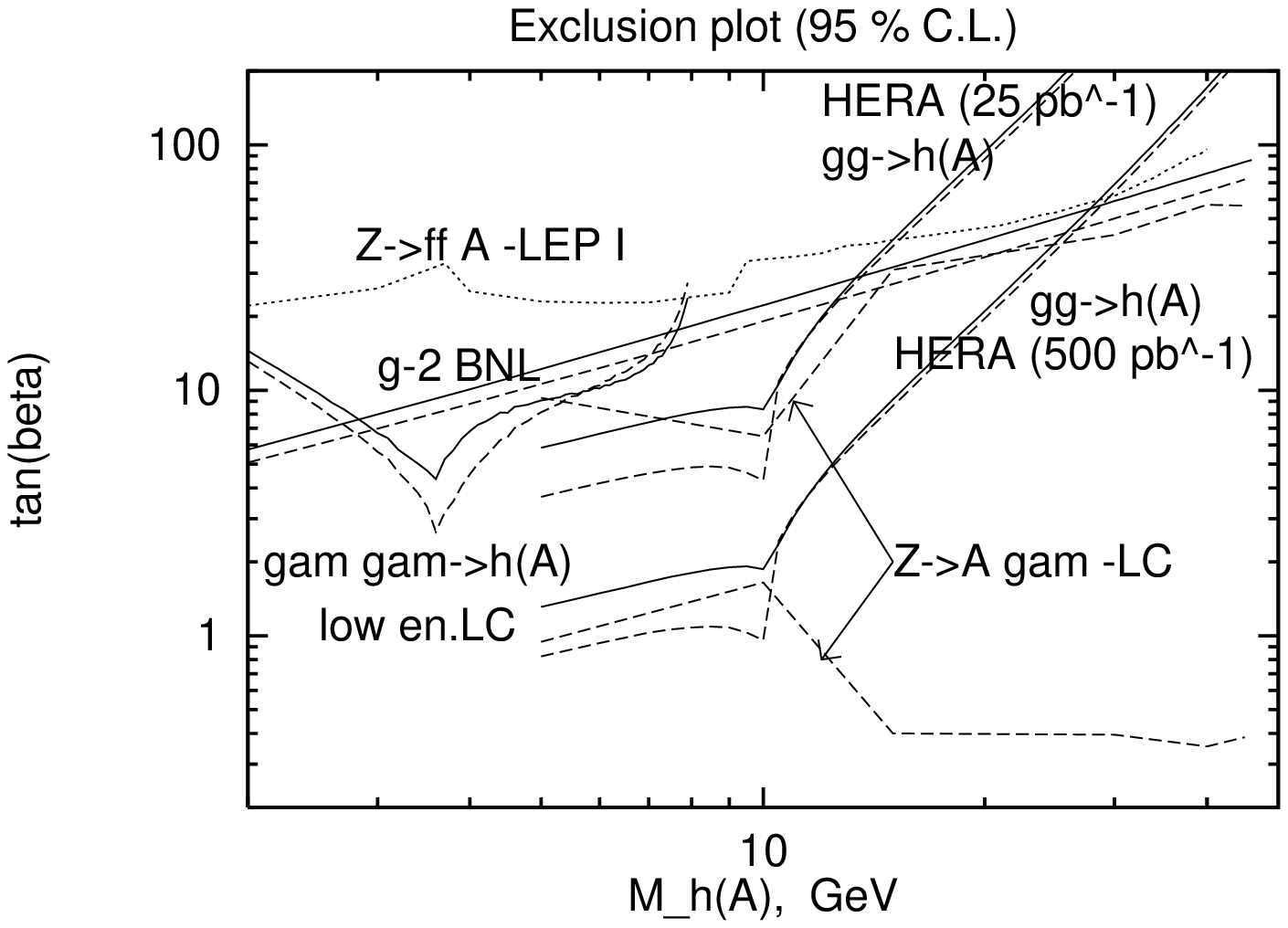,width=120mm}}
\end{center}
\captive{
 The potential of the future data on $g-2$ for muon \cite{g2}, 
the HERA measurement
(the integrated luminosity 25 and 500 $pb^{-1}$),
and the Linear Collider running at low energy $\sqrt s_{ee}$=10 GeV
(with 10$fb^{-1}$)
 and at $Z$-peak (with 20 $fb^{-1}$). For the reference the results 
from the Yukawa process $Z\ra {\bar f}f A$ at LEP I are shown (see also Fig.3).
From \cite{lcmkz}. 
\label{figure}}
\end{figure}
\subsection{The ``Z-factory'' at Linear Colliders}
This process can be studied with higher precision at the planned 
high luminosity ``Z-factory'' at Linear Colliders \cite{ron,lczhag,lcmkz}.
In Fig. 4 the prediction  based on the $Z\ra A + \gamma$,
assuming the integrated luminosity 20 fb$^{-1}$, is shown \cite{lcmkz}.
The $Zh\gamma$ and $ZA\gamma$ couplings can be tested also in
$e \gamma$ option of LC, see Sec.5. 

\section{Structure of photon and a search for a non-minimal Higgs bosons 
at HERA}
The interesting opportunity to  look for  light 
neutral Higgs bosons in the framework of 2HDM 
is due to the photoproduction processes 
at $ep$ HERA collider \cite{gpr} 
(see also \cite{ss}, where the
SM Higgs boson production in processes involving 
the partonic content of the photon was studied). 
Here the   Higgs boson production
 with mass below 40-50 GeV
is dominated by subprocesses 
due to the partonic content of the photon \cite{bk,mk}. 
In particular the process
\be 
g^{\gamma} g^{p} \ra h (A),
\ee
with subsequent decay into $\tau$ pairs was studied in detail \cite{bk}.
We found that for this channel and for the tagged electron
one can, at least in principle,  get rid of a serious background
due to $\gamma g^p \ra \tau^+ \tau^-$. 
On the other hand the   $b {\bar b}$ final state
looks very difficult. 
The expected exclusion for the $\tan \beta$ versus Higgs boson mass 
based on the $\tau$ channel is presented in Fig. 4.
The potential  of the HERA collider to search for a light Higgs boson
is larger than it follows from this plot, as in addition there
are other  contributions due to subprocesses, with and without 
 the partonic content of the photon (like $\gamma b^p$, 
${\bar b}^{\gamma} b^p$ etc.).

Note that the light Higgs bosons from 2HDM 
can be also produced in
$\gamma \gamma$ fusion  in eA collision at HERA \cite{kl}.

\section{$\gamma e$ and $\gamma \gamma$ Linear Colliders} 
The Linear Colliders running as $\gamma e$ and $\gamma \gamma$
 Linear Colliders \cite{telnov,pcilya}, with  high energy photon beams
obtained in the back Compton scattering on a laser light, 
offer excellent probe of the $Zh\gamma$, $ZA\gamma$ and 
$\gamma \gamma h $ or $\gamma \gamma A$ couplings.
All these  couplings  are of great importance for testing
the structure of the Standard Model and of the MSSM or 2HDM
\cite{bbb,mele,pcilya1,lcegama,lowlc}.

 The very low energy $\gamma \gamma$ collider
has been suggested some time ago as a test machine for the NLC \cite{lowtest}.
The potential of such collider 
 in searching for a light neutral Higgs boson in 2HDM 
was studied in \cite{lowlc}.
The results, based on the $h(A)$ decays into muons,
 are plotted in Fig. 4.   
The work for a light neutral Higgs bosons production
in 2HDM at the higher energy photon colliders is in progress.

\section{Summary and outlook}
The processes involving photons play an important role in constraining the
Higgs sector of the Standard Model  and of its extensions. 
The potential of the future Linear Colliders, including Photon Colliders,
in testing the  basic structure of the theory is large.

These colliders have also an unique potential 
 in getting insight in the structure of photon at much higher energy
than present experiments. 
The option $e \gamma$ has an additional advantage -
it allows to test the structure of really {\sl real} photon.  
The reconstruction of  true kinematical variables,
a source of large errors in the present (and presumably future)
experiments measuring the structure
of photon in $e^+e^-$ colliders (see eg. \cite{survey}),
 should be straightforward here. 
The perfect knowledge of the structure of photon is needed not only
to  test the strong interaction sector of the Standard Model \cite{zerwas}
but also to control the background for the New Physics.

\vskip 1 cm
The author is indebted to the Organizers for  interesting and fruitful
Workshop. She is grateful to J. \.Zochowski, P. M\"{a}ttig 
for a nice collaboration and to M. Staszel
and Z. Ajduk for their help in preparation of this contribution.
\noindent
Partly supported by \kbn.

\end{document}